\documentclass[acmtrans,authorversion,nonacm]{acmart}

\usepackage{amsmath,amsfonts}
\usepackage{algorithmic}
\usepackage{graphicx}
\usepackage{textcomp}
\usepackage{xcolor}
\usepackage{url}
\usepackage{microtype}
    
\usepackage{xcolor}
\usepackage{xspace}

\newcommand{\etal}{\hbox{\emph{et al.}}\xspace}
\newcommand{\eg}{\hbox{\emph{e.g.,}}\xspace}
\newcommand{\ie}{\hbox{\emph{i.e.}}\xspace}

\newcommand{\NA}{---}

\PackageWarning{TODO:}{X}
\PackageWarning{TODO:}{X\%}

\usepackage{tcolorbox}

\usepackage{tcolorbox}

\newcommand{\cHRev}{\emph{cHRev}\xspace}

\newcommand{\whodo}{\emph{WhoDo}\xspace}

\newcommand{\Sentinel}{\texttt{RevRecV1}\xspace}
\newcommand{\MetaRec}{\texttt{RevRecV2}\xspace}
\newcommand{\MetaRecWL}{\texttt{RevRecWL}\xspace}

\newcommand{\BystanderRnd}{\texttt{Bystander\allowbreak RecRnd}\xspace}

\newcommand{\Meta}{Meta\xspace}
\newcommand{\timeInReview}{TimeInReview\xspace}
\newcommand{\eyeballTime}{TimeSpent\xspace}
\newcommand{\Phabricator}{Phabricator\xspace}

\newlength{\imagewidth}
\begin{document}

\title{Improving Code Reviewer Recommendation: Accuracy, Latency, Workload, and Bystanders}

\author{Peter C. Rigby}
\orcid{0000-0003-1137-4297}
\affiliation{%
  \institution{Meta}
  \country{USA}
}
\affiliation{%
  \institution{Concordia University}
  \country{Canada}
}
\email{pcr@meta.com}

\author{Seth Rogers}
\orcid{0009-0003-1587-0578}
\affiliation{%
  \institution{Meta}
  \city{}
  \country{USA}
}
\email{sethrogers@meta.com}

\author{Sadruddin Saleem}
\affiliation{%
  \institution{Meta}
  \city{}
  \country{USA}
}
\email{sadruddin@meta.com}

\author{Parth Suresh} 
\orcid{0000-0002-0087-8759}
\affiliation{%
  \institution{Meta}
  \city{}
  \country{USA}
}
\email{parthsuresh@meta.com}

\author{Daniel Suskin} \authornote{This work was done while Suskin was at Meta}
\affiliation{%
  \institution{Meta}
  \city{}
  \country{USA}
}
\email{suskin@meta.com}

\author{Patrick Riggs}
\orcid{0009-0008-9549-8695}
\affiliation{%
  \institution{Meta}
  \city{}
  \country{USA}
}
\email{riggspc@meta.com}

\author{Chandra Maddila}
\affiliation{%
  \institution{Meta}
  \city{}
  \country{USA}
}
\email{cmaddila@meta.com}

\author{Nachiappan Nagappan}
\orcid{0000-0003-1358-4124}
\affiliation{%
  \institution{Meta}
  \city{}
  \country{USA}
}
\email{nnachi@meta.com}

\author{Audris Mockus}
\orcid{0000-0002-7987-7598}
\affiliation{%
  \institution{Meta}
  \city{}
  \country{USA}
  }
  \affiliation{%
  \institution{University of Tennessee, Knoxville}
  \city{}
  \country{USA}
}
\email{audris@meta.com}

\renewcommand{\shortauthors}{P.C. Rigby, S. Rogers, S. Saleem, P. Suresh, D. Suskin, P. Riggs, C. Maddila, N. Nagappan, and A. Mockus}

\begin{abstract}
{\bf Aim.} The code review team at \Meta is continuously improving the code review process. In this work, we report on three randomized controlled experimental trials to improve code reviewer recommendation.

{\bf Method.} To evaluate the recommenders, we conduct three A/B tests which are a type of randomized controlled experimental trial. The unit is either the code diff (\Meta's term for a pull-request) or all the diffs that an author creates during the experimental period. We set goal metrics, \ie those we expect to improve, and guardrail metrics, those that we do not want to negatively impact, \ie analogous to safety metrics in medical trials. We test the outcomes using a t-test, Wilcoxon test, or Fisher test depending on the type of data. 

{\bf Expt 1.} We developed a new recommender, \MetaRec, based on features that had been successfully used in the literature and that could be calculated with low latency. In an A/B test on 82k diffs in Spring of 2022, we found that the new recommender was more accurate and had lower latency. The new recommender did not impact the amount of time a diff was under review. The results allowed us to roll-out the recommender in Summer of 2022 to all of \Meta.

{\bf Expt 2.} Reviewer workload is not evenly distributed, our goal was to reduce the workload of top reviewers. Based on the literature, and using historical data, we conducted backtests to determine the best measure of reviewer workload. We then ran an A/B test on 28k diff authors in Winter 2023 on a workload balanced recommender, \MetaRecWL. Our A/B test led to mixed results. When a low workload reviewer had reasonable expertise, authors selected them, however, the top recommended low workload reviewer was often not selected. There was no impact on our guardrail metrics of the amount of time to perform a review. This workload balancing replaced the recommender from the first experiment as the recommender in production at \Meta. 

{\bf Expt 3.} Engineers at \Meta often select a team rather than an individual reviewer to review a diff. We suspected the bystander effect might be slowing down reviews of these diffs because no single individual was assigned the review. On diffs that only had a team assigned, we randomly selected one of the top three recommended reviewers to review the diff with \BystanderRnd. We conducted an A/B test on 12.5k authors in Spring 2023 and found a large decrease in the amount of time it took for diffs to be reviewed. We did not find that reviewers rushed reviews. The results were strong enough to roll this recommender out to all diffs that only have a team assigned for review. 

{\bf Implications.} Aside from the direct findings from our work, our findings suggest there can be a discrepancy between historical back-testing and A/B test experimental findings, and that more A/B tests are necessary to test recommenders in production. Outcome measures beyond accuracy are important. This is especially true in understanding how recommenders change a reviewer's workload. We also see that the latency in displaying a recommendation can have a large impact on how often authors select recommendations making the reporting of latency an important metric for future work.  

\end{abstract}


\maketitle

\section{Introduction}

Code review is a critical activity in the software development lifecycle that ensures that at least one other experienced developer examines the code for defects and other issues. Code review has evolved from the inspection of massive completed artifacts~\cite{Fagan1976IBM} to the continuous code review of the difference between the changed files in the form of pull-requests, patchsets, and diffs~\cite{rigby2013convergent,bacchelli2013expectations}. A key part of the success of code review is the reviewer who both learns from the change and also suggests improvements. Finding an optimal code reviewer in OSS and a large organization is a complex task, hence substantial research has been devoted to code review recommenders which are tools that suggest who should review the code. At \Meta the reviewers or review teams are recommended to the author who selects the best person or team. The author can also assign reviewers who are outside of the recommendations. Much of that literature uses the assumption that the actual code reviewer was an optimal choice to evaluate the accuracy of the recommenders on historic data. Instead, we run A/B tests of newly developed recommenders to assess if the modifications made to recommenders have the desired effects, goals metrics, and no undesired side-effects, guardrails metrics.   

At \Meta, we are constantly evolving our code review system. In this work, we  describe a series of experimentally validated improvements that we made to our code reviewer recommenders. Our starting point was developed in 2018 and we refer to it as \Sentinel. To address latency and accuracy issues with \Sentinel, we developed \MetaRec that is based on a literature review of successfully used features. Building on \MetaRec, we use measures of reviewer workload to spread review work and develop \MetaRecWL. Finally, we combat the bystander effect when a group is assigned to do a review with \BystanderRnd.
%
One of our major contributions is to experiment with our recommenders in production and to conduct randomized controlled experimental trials (A/B test) on each recommender across hundreds of thousands of reviews and 10's of thousands of engineers. We provide an overview of the motivation for each recommender and briefly summarize the results.

\textbf{Expt 1. Accuracy and latency improvements with \MetaRec.} On the code review team's feedback group, some engineers complained that they were being incorrectly recommended for reviews. When we investigated the current recommender, \Sentinel, we found it was purely based on who had last modified the lines in the current diff (\ie blame lines), and it would suggest reviewers who had moved onto other projects because they had edited the files in the past. We reviewed the literature and developed a recommender that took the most promising features into account. 

\textit{Result summary:} 
In an A/B test, our recommended reviewers were more likely to be the ones actually conducting the review with a 14.19 percentage point improvement over the previous model. The recommendations were also displayed more quickly with a 14 times reduction in latency at the 90th percentile. The recommendations were selected by authors 27\% more often with the new model. We did not see any statistically significant regressions in our guardrail metrics including the review cycle time and the reviewer viewing time. In Summer 2022, \MetaRec was rolled out to 100\% of engineers.

\textbf{Expt 2. Reviewer workload balancing with \MetaRecWL.} The number of reviews that each engineer does is known to be skewed~\cite{rigby2014peer,Hajari2022Thesis,Al-Zubaidi2020PROMISE}. At Microsoft~\cite{asthana2019whodo} and Ericsson~\cite{Strand2020IEricsson}, they implemented a recommender that weighted the expertise based recommender score by the workload of the candidate reviewer. We replicate these works and experiment with novel workload measures, including the number of scheduled meetings to understand if workload balancing is effective.

\textit{Result summary:}
When a reasonable candidate with lower workload was available, we saw a large reduction in workload. When an inappropriate candidate was recommended, simply because they had low workload, the author looked down the list for a more experienced developer, and we saw a drop in the top ranked candidate being selected. We did not see any statistically significant change in our guardrails: review cycle time or latency. In Winter of 2023, \MetaRecWL was rolled out to 100\% of engineers.

\textbf{Expt 3. Reducing the bystander effect with \BystanderRnd.} The bystander effect occurs when nobody is explicitly assigned responsibility for a situation and no individual takes action or they are delayed in taking action~\cite{fischer2011bystander}. Beyond noting that it may be an issue in code review~\cite{Meneely2014}, we did not find any research in the software engineering literature. At \Meta teams and groups of developers can be assigned a review without any explicit individual. To reduce the bystander effect, we use our recommenders to explicitly assign a reviewer. 

\textit{Result summary:} 
At \Meta, we randomly assigned one of the top three recommended reviewers to review the diff and found a drop of -11.6\% in \timeInReview with no regressions in our guardrail metrics. In Spring of 2023, we rolled \BystanderRnd out to 100\% of diffs that only had a team assigned to do the review. 

The remainder of this paper is structured as follows. In Section~\ref{secBackground}, we describe \Meta and its code review process. In Section~\ref{secLiterature}, we review the literature to identify the features that have been successfully used to recommend reviewers, and the design of our recommenders. In Section~\ref{sec:methodology}, we describe our experimentation methodology. In Section~\ref{secResultRQ1}, we update the recommender at \Meta to include these features and experiment with a new model in production and report results for Expt 1. In Section~\ref{secWLResults}, we weight the recommender by the workload of the candidate reviewers in an attempt to balance workload and report results for Expt 2. In Section~\ref{secBystanderResults}, for Expt 3, we reduce the impact of bystander effect when no individual is assigned a review by randomly assigning one of the recommended reviewers. In the final sections of the paper, we discuss threats to validity in Section~\ref{secThreats}, position our work in the literature in Section~\ref{secLitDiscuss}, and conclude the paper in Section~\ref{secConclusion}.

\section{Background and Code Review Process}
\label{secBackground}

\begin{figure*}
\centering
\includegraphics[width=2\imagewidth]{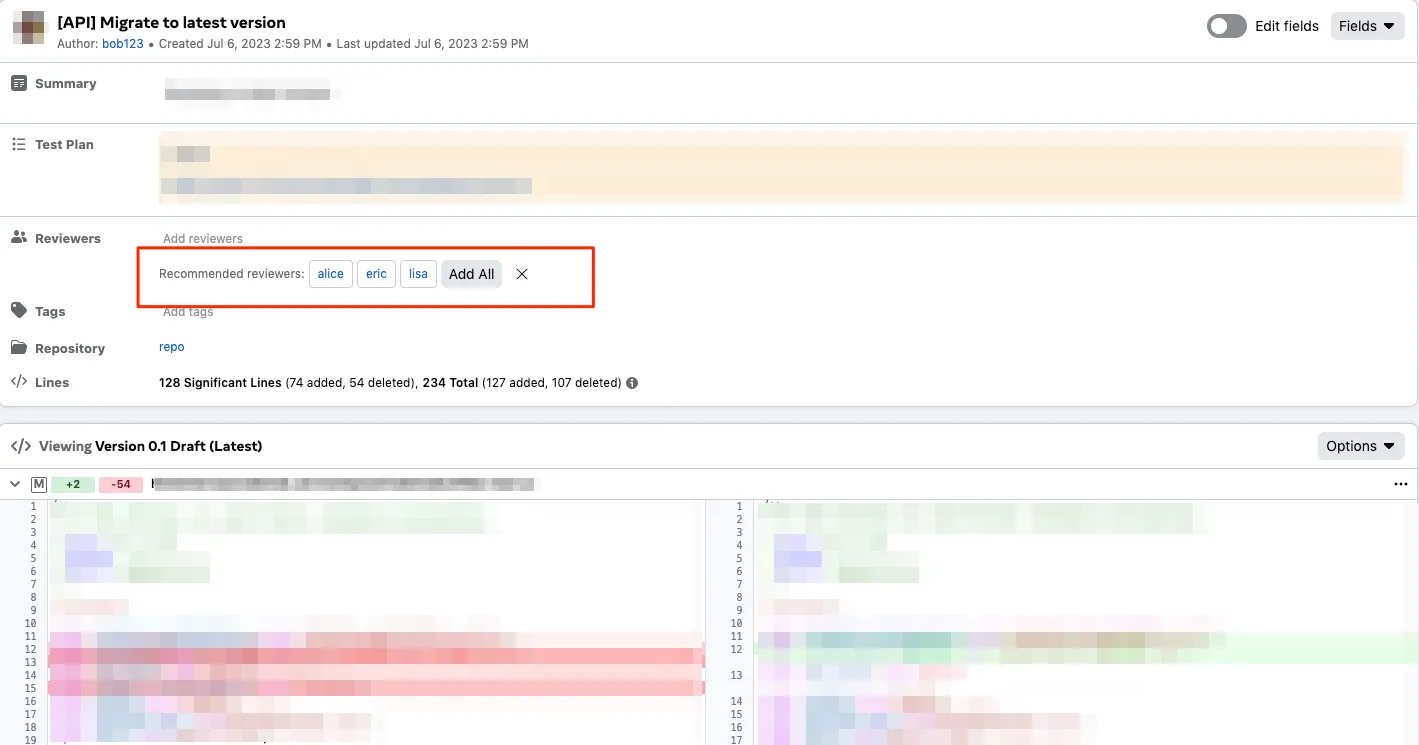}
\caption{A code change, \ie diff, under review at \Meta. The recommended candidates for the review are clickable by the author and ordered left to right with the top ranked candidate on the left (see the red box in the figure). We have anonymized the diff and used mock names.}
\label{figRevRecommendation}
\end{figure*}

\Meta is a large online services company that works across the spectrum in the communications, retail, and entertainment industries. \Meta has tens of thousands of employees with offices spread across the globe (North America, Europe, Middle East and Asia). \Meta has its own dedicated infrastructure teams where the tools used are a mix of commercial and in-house developed systems. 
We study code reviews from the projects using \Phabricator\footnote{Phabricator is an open source project available at \url{https://www.phacility.com/phabricator/}} which includes user facing social network products and virtual and augmented reality projects as well as software engineering infrastructure, such as calendar, tasks, and release engineering tooling. 

\subsection{Code review process} 

Code review at \Meta follows a similar structure to the contemporary practices used at Google~\cite{sadowski2018modern} and Microsoft~\cite{bacchelli2013expectations}, and we briefly summarize the practice here. Figure~\ref{figRevRecommendation} provides an example of a diff review in Phabricator with portions redacted and using mock names. The author fills in a number of fields including the title and a detailed summary of the diff change. A test plan is provided that details which steps the author took to validate their changes, and provides reviewers a guide to test the change themselves. Line-by-line highlighting is used to show additions and removals, and inline comments are used to indicate any compilation or linting issues. The diff will also show the output of tests that were required to run based on the files included in the change.

To start a code review, the authors must select one or more reviewers and/or reviewer groups. The author can explicitly type in names, or select recommended reviewers. The latency is a key consideration for a recommender because slow recommendations will not appear in time to be selected by the author. Once selected, the reviewer or a group of reviewers are notified via email or chat. The reviewers provide feedback, ask questions, and make suggestions, which should be addressed by the author. There may be multiple rounds of feedback and changes before a reviewer accepts/rejects the diff. \Meta uses a single track/monorepo development model, so once the diff is accepted by at least one reviewer it will be integrated for further testing and ultimately ``shipped`` into production.

For this paper, we focus on the reviewer recommendation experience. We see the recommended reviewers highlighted with a red box in Figure~\ref{figRevRecommendation}. We adapt the approaches behind the recommendations, but leave the interface consistent and unchanged.

\subsection{Reviewer Recommenders at \Meta}

The reviewer recommender used in production at \Meta before we began our work was developed in 2018. For this study it will be our starting comparison point and we will refer to it as \Sentinel. \Sentinel considered the following attributes: the blame lines \ie the person who has authored the most lines in a file is an expert, authorship frequency \ie the person who has made the most changes to a file is an expert, and review frequency \ie the number of times a reviewer has accepted the changes to a file. \Sentinel is trained using a genetic algorithm, an ML approach popular in 2018.
While the recommender has been effective and makes thousands of recommendations every day, the code review team identified a number of issues with it. First, many developers complain about incorrect recommendations on the internal feedback group. Second, blame is a computationally expensive command that can delay the loading of the code review page for a diff. For these reasons, the code review team decided to test alternative strategies for recommending reviewers to improve accuracy and reduce latency.

\section{Literature, Features, and Recommender Design}
\label{secLiterature}

We reviewed the literature to identify the reviewer recommendation approaches and model features that were most successful in existing recommenders. Like Microsoft~\cite{asthana2019whodo,Zanjani2016TSE} and Ericsson~\cite{Strand2020IEricsson}, we cannot use features in production that would add latency to our recommender. We also found that very few recommenders that are studied in the literature report latency and even fewer are actually deployed into production. In this section, we update Al-Zubaidi \etal's~\cite{Al-Zubaidi2020PROMISE} literature review and use their categories: code ownership, reviewing experience, review participation rate, and familiarity with the author. Table~\ref{tableRecFeatures} shows the features we selected for \MetaRec. In Section~\ref{secDesignWorkload}, we discuss how we weight \MetaRec by different types of workload to create \MetaRecWL.

\subsection*{Code Ownership} 

The influence of code ownership on code quality has been extensively investigated in the literature\cite{bird2011don,rahman2011ownership,foucault2015impact,thongtanunam2016revisiting} with recommenders to identify expert developers for a particular part of code or task~\cite{mockus2002expertise}. Researchers have used a wide range of granularity, from lines~\cite{fritz2007does, girba2005developers,rahman2011ownership} to modules~\cite{bird2011don}, to estimate ownership of developers. 
Studies on code review find that code owners are usually selected to review changes~\cite{sadowski2018modern,bacchelli2013expectations,greiler2016code}.
In this work, we quantify ownership as the number of files under review that a candidate reviewer has modified in past diffs.

\subsection*{Reviewing Experience}

Thongtanunam \etal~\cite{thongtanunam2016revisiting} recommend reviewers based on the files that an engineer has reviewed. Intuitively, engineers who have reviewed the files in the diff in the past, will be good candidate reviewers. Zanjani \etal~\cite{Zanjani2016TSE} builds upon this notion of reviewer expertise by adding the number of comments a candidate reviewer has made on the files under review. 
In this work, we count the number of comments as well as other reviewer actions such as accepting or rejecting the diff.

\subsection*{Familiarity with Patch Author}

Moving beyond code ownership, reviewer recommenders have begun to consider relationships among the reviewer and the author~\cite{bosu2016process}. The simplest metric is the number of times a reviewer has commented on an author's diffs in the past. More sophisticated usages include social relationship graphs~\cite{yu2016reviewer}. 
With 10's of thousands of developers, and thousands of diffs every day, social network graphs are not performant. In this work, we suggest a novel usage of how long a reviewer looks at an author's diffs: the time a reviewer spent looking at the diffs the author wrote in the code review tool. While \Meta has sophisticated tracking of how long an engineer is looking at a webpage and uses this to capture how long a reviewer looks at a review, other projects can easily add this to the review tool. For example, the SmartBear code review tool has been tracking this information since 2006~\cite{Cohen2006} to identify reviewers that looked at the change for too short a period of time given the size of the diff, \ie ``rubberstamping". The web abounds with code to monitor time spent on a webpage.\footnote{A JavaScript library that tracks time spent on a page \url{https://stackoverflow.com/questions/4667068/how-to-measure-a-time-spent-on-a-page}}

\subsection*{Review Participation Rate}

On open source software projects, the number of reviews that a developer has been assigned and participated in is a strong predictor of whether they will perform a review~\cite{Ruangwan2019EMSE,rigby2014peer}. We measure this participation rate in two ways, the number of diffs that a reviewer was assigned and the number of diffs that a reviewer resigned from. These two measures capture how likely a reviewer is to participate in reviewing. 

\subsection*{Workload}

Recent works have incorporated the reviewer workload into reviewer recommendation. Early works quantified the relationship between time waiting for a review and review workload. Baysal \etal~\cite{Baysal2013WCRE} studied WebKit to understand the impact of metrics on review time. They found that when reviewers had longer queues of waiting reviews, reviews tended to take longer. They also found that more active reviewers had quicker response times than less active reviewers. Studying four open source projects, Ruangwan \etal~\cite{Ruangwan2019EMSE} found prior participation and prior experience were the best predictor of whether a reviewer will respond to a review request. The familiarity of the author and reviewer had a minor impact on participation.

Microsoft weighted their code reviewer recommender, \cHRev, by the number of open reviews a candidate, \whodo, had and found between a 14\% and 21\% improvement in the time taken for reviews depending on the project~\cite{asthana2019whodo}. In contrast, Strand ~\etal~\cite{Strand2020IEricsson} evaluates a workload aware recommender at Ericsson. In a backtest they found that a recommender aware of the file change history and the level of recent workload activity was able to make recommendations with an MRR of 0.37 or an average rank of 2.7. Despite these reasonable recommendations, after manually evaluating the performance of the recommender on 47 changes they determined that it did not reduce lead time or workload for code review and they decided to not roll-out the recommender.

\begin{table*}
\caption{The Features for \MetaRec and their importance in the model deployed in Section~\ref{secResultRQ1}. We only include features that can be calculated with low latency, we use the feature categories suggested by Al-Zubaidi \etal~\cite{Al-Zubaidi2020PROMISE}. Workload is only part of \MetaRecWL and is not included in this table.} 
\begin{center}

\begin{tabular}{ p{.40\imagewidth}| p{.20\imagewidth} | p{1\imagewidth} || r}

{\bf Category}	&	{\bf Feature}	&	{\bf Description}	&	{\bf Importance}	\\ \hline
Familiarity with Author	&	time spent	&	The time the candidate spent in the diff review tool looking at the author’s diffs in the past 90 days	&	33.89\%	\\ \hline
Familiarity with Author	&	explicit time spent	&	The time the candidate spent in the diff review tool looking at the author's diffs when they were explicitly assigned as a reviewer in the past 90 days	&	6.69\%	\\ \hline
Review Participation Rate	&	assigned reviews	&	Number of diffs the candidate was assigned as reviewer. At \Meta reviewers are assigned to many diff either directly or as part of a team or review group.	&	31.28\%	\\ \hline
Review Participation Rate	&	resigned diffs	&	Number of diffs the candidate resigned from. When a reviewer does not feel qualified to review a diff they can resign from that review.	&	0.00\%	\\ \hline
Reviewing Experience	&	rejected diffs	&	Number of diffs the candidate rejected	&	11.56\%	\\ \hline
Reviewing Experience	&	reviewer comments	&	Number of diffs the candidate commented or acted on as a reviewer 	&	10.10\%	\\ \hline
Reviewing Experience	&	total comments	&	Number of diffs the candidate commented on	&	1.57\%	\\ \hline
Code Ownership	&	authored diffs	&	Number of diffs the candidate authored in the past which touched any of the files in the diff.	&	4.91\%	\\ 

\end{tabular}
\label{tableRecFeatures}
\end{center}
\end{table*}

\subsection{Design and Feature Importance for \MetaRec}

Prior works have mostly ranked reviewers based on a set of features and rules, \eg~\cite{asthana2019whodo,Mirsaeedi2020ICSE}. Since reviewer recommendation is a ranking problem, we select learn-to-rank approach~\cite{Burges2010LTR}, LambdaMART. A pairwise comparison of each reviewer is made and this ordering is then used to determine the final order. In our case we use MART (Multiple Additive Regression Trees) as the function to determine the ordering of reviewers. 
The LambdaMART is the boosted tree version of LambdaRank, which is based on
RankNet. All have proven to be very successful algorithms for solving real world ranking problems. 
An approximate description of how it works would take more space than this paper and is presented in~\cite{Burges2010LTR}.  
At \Meta there is infrastructural support for LambdaMART models at scale using measures collecting data from internal databases on code review and coding.

{\bf Feature importance.} In decision trees, feature importance is determined by how much each feature contributes to reducing the uncertainty in
the target variable by the amount of reduction in the entropy that is
achieved by splitting on a particular feature. Since LambdaMART uses boosted regression trees, the average reduction in entropy is calculated over 
the trees.  The single-tree entropy is calculated by iterating over the non-leaf nodes of the tree and obtaining the weighted reduction in node entropy from the split at that node.

To train \MetaRec, we used the features shown in Table~\ref{tableRecFeatures}. The features were selected by the engineering team manually by using their domain knowledge of code review practices and their understanding of what metrics could be deployed in production. Specifically, they decided which features are most likely to contribute to the model's prediction and then selected a subset that would would not slow down the recommender. Some features were included because the engineering team felt that they were inherently interesting. For example, although resigning from a diff has little predictive power, the engineers felt that a resignation demonstrated a clear example of a wrong reviewer suggestion. Since the number of features was small and prediction performance was stable, additional feature engineering was not conducted in this study especially since experiments showed substantial improvements over the baseline model.
We trained the model on over 4.3 million instances of these features for reviewers. We evaluated the model on 1.1 million recommended candidates. 

In Table~\ref{tableRecFeatures}, we show the feature importance across the 1.1 million diffs. The most important features that contribute to this accuracy improvement relate to the familiarity of the author with the review (time spent on the author's prior diffs, 33.89\%) and the reviewer participation rate (assigned reviews 31.28\%). In contrast, code and review ownership are still useful but much less important. The number of files in the current diff that the candidate reviewer has authored in the past accounts for only 4.91\% (authored diffs). The number of diffs that a reviewer has acted on and rejected have an importance of 10.10\% (reviewer comments) and 11.56\% (rejected diffs). The time spent when assigned a diff (explicit time spent) accounts for 6.69\%. The total number of comments and number of resigned diffs have low importance with 1.57\% and 0.00\%, respectively. Our recommender clearly show that being a more active reviewer and having reviewed the author's diff frequently in the past are much stronger predictors of who will actually perform a review than code ownership.

\subsection{Design for \MetaRecWL} 
\label{secDesignWorkload}

We design \MetaRecWL, which adds Workload (WL) to \MetaRec. Our goal is to identify the top expert developer with the {\it lowest current workload}. To this end, we adapt the approach taken at Microsoft~\cite{asthana2019whodo} that balances the workload of reviewers by weighting the score by the current workload of candidate reviewers for a diff review, $r$:

\begin{equation}
\text{\text{Load}(r)} =  e^{\theta.Workload} 
\end{equation}

$\theta$ is a parameter between 0 and 1 to control the amount of
load balancing.

At Microsoft~\cite{asthana2019whodo}, they found that initial feedback from developers indicated that reviewers from other projects complained that they were being incorrectly recommended. We made three compromises to convince the engineering team to run the experiment: (1) we set $\theta = 0.1$ which reduces the influence of workload favoring expertise, (2) we only re-rank the top five suggestions from \MetaRec, and (3) instead of assigning the top ranked reviewer, we allow the author to select one of the top five reviewers. These latter two are important guarantees, first, people with low workload, but little experience will not be inadvertently added to the list eliminating the issue that Microsoft experienced. Second, the Top5 accuracy of \MetaRecWL = \MetaRec. Since the top five developers are displayed in \Phabricator, this means in the most extreme case, the top developer from \MetaRec will be displayed as the 5th developer, if they have high workload, but the same reviewers will be displayed only the order will change. 

For a diff review, $r$, \MetaRecWL final scoring function is
\begin{equation}
\text{\MetaRecWL}(r) = \frac{\text{\MetaRec}(r)}{\text{Load}(r)}
\end{equation}

Instead of choosing a single measure of workload, we experiment with three types of workload weights, two are novel: the number of upcoming meetings in the next seven days, the reviewer activity in the last seven days~\cite{asthana2019whodo}, and time spent reviewing in the last seven days. We calculate each over a seven day period because developer patterns on a particular day tend to vary, but the entire work week tends to be more predictable~\cite{LChen2022FSE-industry}. In Section~\ref{secWLResults}, we present our historical analysis to determine the best workload measure and our A/B test to determine if we can use \MetaRecWL in production. 

\subsection{Design for {\tt RecBystander}}

The bystander effect is a well known psychological problem that when a group of people are present, it is less likely that any individual will take action~\cite{fischer2011bystander}. At \Meta, developers work in a monorepo, so there are times when the author does not know which engineer to assign the review to. In other cases, the author might not mind and anyone on a particular team can do the review. We decided to randomly assign one of the top three reviewers from \MetaRec to the review. The results for \BystanderRnd are presented in Section~\ref{secBystanderResults}.

\section{Experimental Method and Outcome Metrics}
\label{sec:methodology}

Industry experiments are rarely reported in software engineering literature with most empirical studies being observational, where past data is analyzed via a regression analysis in an attempt to suggest causal relationships. In observational studies the treatment may be associated with some unobserved property of the subjects or tasks that affects the outcome \ie allocation bias. Furthermore, the code recommender literature (\eg \cite{thongtanunam2015should}) assumes that the actual reviewer observed in the historic data is the gold standard (correct answer) for any proposed recommender. This ignores the possibility that the actual reviewer may not always represent the optimal choice and that the developer would not be affected by the recommendations they might be shown if a different recommender were to be used. This ``backtesting'' approach consists of calculating recommendations based on data predating the reviewer assignment and then checking if the actual reviewer is from among the TopN recommended reviewers. 

A more reliable way to establish treatment effects is to run a randomized controlled experimental trial~\cite{jadad2007randomized}. In software engineering these trials are usually called A/B tests (see, for example,~\cite{juristo2013basics}, for the discussion of various types of experimentation in software engineering).  Per definitions in~\cite{cook2002experimental}: an experiment is a study in which an intervention is deliberately introduced to observe its effects and in randomized experiments units are assigned to receive the treatment or an alternative condition by a random process. Industry experiments in software engineering are rare~\cite{sjoberg2005survey,juristo2016experiences} for a number of reasons. For example, experiments involving creation of  systems or features in parallel (see, e.g., ~\cite{anda2008variability}) are prohibitively expensive. Involving human subjects also requires addressing numerous ethical and privacy concerns. Some of the early experiments in software engineering did, however, involve code reviews~\cite{porter1998understanding}. 

Specifically, we are running experiments with a randomized intervention where each unit is in an A or B 
 condition. The randomization assures that observed (and unobserved) context will be evenly distributed under the A or B conditions.
 In our case, each experiment includes an older recommender, condition A or baseline condition, and a new recommender, condition B. Like in clinical trials of drugs, for example, a drug should have an intended effect, such as reducing inflammation. Similarly, the new recommender was designed with a goal in mind, e.g. workload balancing. The drug, even if it is highly effective at reducing inflammation, may have undesired side effects, such as higher mortality and such side effects are recorded in clinical trials. If the side effects are stronger or more frequent in the experiment than in the baseline condition, the trial may need to be stopped. Similarly, we do not want to negatively impact our code review process, so we have guardrail metrics like how long a reviewer spends looking at a diff because we do not want people to start rubberstamping and spending less time on review because of our change in recommendations. If we see regressions in these guardrail metrics the experiment can be stopped early or will not be deployed to the entire company.

A/B test requires that a treatment (specific recommender) be assigned to subjects (engineers in parallel design or individual diffs in crossover design) at random with another option being the baseline (the existing recommender). 
The main advantage of random assignment is the removal of the allocation bias, i.e., it ensures that the distribution of the potential confounders are independent of the treatment. Also, the researcher can not assign treatments (and, thus, can not steer certain reviewers to certain treatments). Since the user interface in each treatment is exactly the same, the engineer is also not aware if they are in the baseline or treatment conditions so they can not intentionally act differently depending on the condition.

In contrast, if we were conducting an observational study, we would need to include numerous control variables, 
for example, the size and complexity of the diff (and of the underlying codebase), author and reviewer expertise, programming language, type of functionality being implemented, time of the year, day of the week, and numerous other factors that may affect 
the outcome and can (or in many contexts can not) be measured. Using randomization helps to control for both observed and 
unobserved variables. Results from A/B tests are, therefore, relatively easy to analyze. Because of the random assignment, on average, the only differences between recommenders are due to recommenders. 

While software engineering experiments are rare, the software industry commonly uses the so-called A/B testing~\cite{kohavi2020trustworthy} to describe the evaluation protocol for, typically, user interface modifications. For example, when a user goes to a web page, they are randomly assigned either a regular version of a service or a modified version. Typically many different trials are ongoing at the same time so the randomization follows a factorial design to ensure that effects of each intervention (and, if possible, some of the interactions) can be estimated from the data. To implement A/B test for reviewer recommender, we ran a one-factor two-treatments controlled experiment as the two recommenders were the only distinction between the treatments. In many such experiments the user may see a visual difference and recognize that they are receiving experimental treatment and act differently, but in this case no such difference was observable, hence the treatment was ``blinded” from the  subject.

To conduct our experiments, we used an internal tool designed to support A/B testing. The tool is programmed to do randomization where, in some experiments, we used parallel design where each developer gets only one recommender, and in others we use a crossover design where each diff gets a randomly chosen recommender so that each developer gets assigned one recommender for some of the diffs and another for other diffs. The tool is then programmed to collect both goal and guardrail metrics and to run during a preset period of time. Once the experiment is complete, it presents an analysis page where average value and a confidence interval (at pval=0.01) are shown for each metric. The tool uses Welch's t-test for unpaired samples to calculate the confidence interval. If appropriate permissions are secured, researchers may access raw collected data for usually 90 days but up to 6 months after the experiment. After that the data is deleted only the summaries (means, sample sizes, and standard deviations are available). 

From the developers' perspective, the review recommendation is a service that, 
when an engineer loads the diff page, gets automatically invoked and provides its recommendations. It is during that invocation that each diff may get a different recommender (or a developer keeps the same algorithm for subsequent diffs in case of parallel design). Once the page is loading, the developer may interact with the recommender (due to computational latency the recommender may not always appear immediately),
or type in reviewer name directly. Hence, the record, in addition to the time of the action and the actual reviewer(s) chosen, also indicates if the recommender was interacted with (clicked on). 

As in any experiments involving human subjects, the experimental design is reviewed to ensure the subjects will not be harmed and the tool used provides strong privacy protections. For example, the researcher does not have automatic access to the data that can be used to identify subjects, the data is provided by the tool in an aggregate form (means and standard deviations for each condition), and UID (user identifiable data) is completely removed after 90 days.   



\subsection{Goal and guardrail Outcomes Metrics}
\label{sec:goalDefinition}

The vast majority of reviewer recommendation work does not release the recommender in production and only evaluates the accuracy on historical data. In contrast, the culture at \Meta uses a sophisticated A/B test of every new feature and change. In Table~\ref{tabExperimentDesign} we show the recommenders that are being compared, the unit and size, and the time period of the experiment. 

The team defines goal and guardrail metrics in advance of the experiment and then an A/B test is used to test the existing feature (the control or A condition) with the new feature (the test or B condition). The goal metrics measure the desired impact of the intervention.  In contrast, guardrail metrics are the outcome metrics we do not want to negatively impact. In both cases, the null hypothesis or H0, is that nothing will change under the intervention. In Table~\ref{tabDeltoid}, we introduce the metrics that we use as outcomes in this paper: TopN accuracy, \timeInReview, \eyeballTime, Latency, Clicks on recommendations, and Workload. Depending on the experiment these outcomes will either be goal metrics or guardrails. For continuous data, we use a t-test, and for count data we use a Fisher test. Since workload is skewed we use a Wilcoxon signed-rank test. We often have multiple hypotheses, so we use a 99\% confidence interval. We also report the actual p-value when the result is not statistically significant. We report the percentage point change rather than effect size measurements like small, medium, and large because engineers prefer percentage points as they can more easily discuss the magnitude and importance of the change in the context of the problem they are solving. 

\begin{table*}
\caption{We conduct three experiments. In each experiment we have a control recommender vs a test recommender. We report the experimental unit and when the experiment was run.} 

\begin{center}

\begin{tabular}{ p{.25\imagewidth} | p{.8\imagewidth} | p{.16\imagewidth} | p{.16\imagewidth} | p{.18\imagewidth} }

{\bf Experiment}	&	{\bf Description}	&	{\bf Expt. 1}	&	{\bf Expt. 2}	&	{\bf Expt. 3}	\\ \hline
Conditions: A vs B	&	We conduct controlled experiments with a control or A condition vs a test or B condition	&	\Sentinel vs \MetaRec	&	\MetaRec vs \MetaRecWL	&	TeamOnly vs \BystanderRnd	\\ \hline 
Experimental Unit	&	The experimental unit is either the diff or the author. In either case, the unit is evenly divided across control and test conditions.	&	82k diffs	&	28k authors	&	12.5k authors	\\ \hline
Experiment Date	&	We cannot report the exact timeframe that we ran the experiment, but we report the season in which each experiment was conducted.	&	Spring 2022	&	Winter 2023	&	Spring 2023	\\

\end{tabular}
\label{tabExperimentDesign}
\end{center}
\end{table*}

\begin{table*}
\caption{For each experiment, we describe each outcome measure, indicate whether it is a Goal measure or a guardrail measure, and describe the statistical test that is appropriate for the type of data.} 
\begin{center}

\begin{tabular}{ p{.25\imagewidth} | p{.8\imagewidth} | p{.16\imagewidth} | p{.16\imagewidth} | p{.18\imagewidth} | p{.1\imagewidth} }

\hline

{\bf Outcome measures}	&	{\bf Description}	&	{\bf Expt. 1}	&	{\bf Expt. 2}	&	{\bf Expt. 3}	&	{\bf Stat Test}	\\ \hline
TopN Accuracy	&	How often does one of the recommended reviewers review the diff? We report Top1 and Top3 accuracy.	&	{\bf Goal}	&	guardrail	&	NA	&	Fisher	\\ \hline
TimeInReview	&	The time when a diff is waiting for review or is being actively reviewed. It excludes the time when the author is reworking the code. We want to make sure that our recommender does not slow down reviewers.	&	guardrail	&	guardrail	&	{\bf Goal}	&	t-test	\\ \hline
\eyeballTime	&	The amount of time reviewers spend examining each diff. This is time spent on the review page for a diff. We want to make sure that reviewers are not rushing through reviews and rubberstamping diffs. 	&	guardrail	&	guardrail	&	guardrail	&	t-test	\\ \hline
Latency	&	The amount of time from when the diff is published until the recommendations are displayed. We monitor the 90th and 99th percentile. The goal is to make sure that most diffs have recommendations produced by the time the diff page has loaded. 	&	{\bf Goal}	&	NA	&	NA	&	none	\\ \hline
Clicks	&	The number of times that an author selects one of the recommended reviewers. We want to know how often the recommendations are selected. 	&	{\bf Goal}	&	guardrail	&	NA	&	Fisher	\\ \hline
Workload	&	We measure the amount of time a candidate reviewer has spent looking at diffs in the last week. The distribution is skewed and our goal is to distribute review workload more evenly. 	&	NA	&	{\bf Goal}	&	NA	&	Wilcoxon signed-rank test	\\

\end{tabular}
\label{tabDeltoid}
\end{center}
\end{table*}

\section{Expt. 1: Accuracy and Latency}
\label{secResultRQ1}


In Section~\ref{secLiterature}, we examined the code reviewer recommendation literature. We use features from each category of measures we identified in the literature~\cite{Al-Zubaidi2020PROMISE}: code ownership, reviewer experience, reviewer participation rate, and familiarity with the author. We also use a novel measure of how long a reviewer has viewed the author's changes in the last 90 days called \eyeballTime. 

We conducted an A/B test experiment to understand if the new model is effective in production. Our experimental unit is the diff under review. Diffs in the control group, A, continued to receive reviewer recommendations from \Sentinel, the existing model in production. Diffs in the test group, B, received reviewer recommendations from our new model. Our experiment ran in late Spring of 2022, on 82k diff which were evenly and randomly split between the two conditions. We use a t-test on continuous data and a Fisher test on count data. Since we have multiple goal metrics we use a 99\% confidence interval, \ie $p < 0.01$.

Our goal metrics, the metrics we wanted to improve, were the Top1 and Top3 recommendation accuracy. TopN accuracy in this context is the fraction of times where at least one person belonging to the TopN predictions has either commented or acted, \eg accepted, the diff. We also monitor when a recommendation is clicked, and we expect to see more accurate recommendations being clicked more often. Before we released our new recommender into production, we calculated the latency. In a dashboard we tracked the slowest diffs, \ie the latency at the 90th and 99th percentiles. Because latency was measured outside of the experiment framework, we were not able to calculate statistical significance on this measure.

Our guardrail metrics, the metrics we did not want to be negatively influenced, where the time diffs were in review (cycle time) and the \eyeballTime that reviewers looked at each diff. We did not want to increase cycle time and we did not want recommended engineers to rubber stamp diffs and spend less time looking at them.  

\subsection*{Experimental results: \Sentinel vs \MetaRec}
\label{secMetaRecExperiment}

For Expt. 1, our goal was to improve the accuracy and latency. We had \Sentinel as the control and \MetaRec as the test. The unit was 82k diffs in the Spring of 2022. As we can see in Table~\ref{tabExpt1Results}, the experiment achieved our goals. We also found that the reduced latency improved the number of clicks on recommendations because they were displayed sooner. We did not want to see reviewers rushing reviews and did not expect to see a drop in how long reviews took.

\begin{table*}[h]
\centering
\caption{Expt. 1 the change in accuracy, clicks, latency, \timeInReview, and \eyeballTime for \Sentinel vs \MetaRec on 82k diffs. Latency measures were calculated in a dashboard and no statistical test was conducted, the magnitude of the change is very large from an engineering perspective. Guardrail metrics are safety checks that we do not expect to see change.}

\begin{tabular}{ l | l | l | c | l }

{\bf Outcome Metric}	&	{\bf Type}	&	{\bf Expectation}	&	{\bf Change}	&	{\bf p-value}	\\ \hline
Top1 Accuracy	&	Goal	&	Increase	&	$+9.14$ pp.	&	$p \ll 0.01$	\\
Top3 Accuracy	&	Goal	&	Increase	&	$+14.19$ pp.	&	$p \ll 0.01$	\\ \hline
Clicks	&	Goal	&	Increase	&	$+27.4\%$	&	$p \ll 0.01$	\\ \hline
Latency p90	&	Goal	&	Decrease	&	$-14x$	&	\NA	\\ 
Latency p99	&	Goal	&	Decrease	&	$-9x$	&	\NA	\\ \hline 
TimeInReview	&	Guardrail	&	No Increase	&	\NA	&	$p = 0.33$	\\ \hline
\eyeballTime	&	Guardrail	&	No Decrease	&	\NA	&	$p = 0.58$	\\

\end{tabular}
\label{tabExpt1Results}
\end{table*}

\begin{table}
\centering
\caption{Improvement in Accuracy  for \MetaRec. The improvement is statistically significant with $p \ll 0.01$.}
\begin{tabular}{ l | r | r | r }

	&	Top-1 	&	Top-3	\\ \hline
\Sentinel	&	43.40	&	59.05	\\ \hline
\MetaRec	&	52.54	&	73.24	\\ \hline \hline
Improvement	&	9.14 points	&	14.19 points	\\ 

\end{tabular}
\label{tabAccuracy}
\end{table}

\begin{table}
\centering
\caption{Improvement in Latency for \MetaRec. We measure how long it takes for the recommender to make a recommendation in production, \ie the latency. We are interested in the longest latency, so we report the 90th and 95th percentile, \ie p90 and p95.}
\begin{tabular}{ l | r | r}

	&	Latency at p90	&	Latency at p99	\\ \hline
\Sentinel (control)	&	4.43 s	&	12.37 s	\\ \hline
\MetaRec (test)	&	300 ms	&	1.26 s	\\ \hline
Improvement	&	14x improvement	&	9x improvement	\\ 

\label{tabLatency}
\end{tabular}
\end{table}

\begin{table}
\centering
\caption{We report the number of diffs that the models were able to create recommendations for and how often the author clicked on the recommended reviewer. The improvement in clicked recommendations is statistically significant with $p \ll 0.01$.}
\begin{tabular}{ l | r | r}

	&	Diff with recs	&	Clicked recommendation	\\ \hline
\Sentinel (control)	&	38,247	&	6,420	\\ \hline 
\MetaRec (test)	&	41,692	&	8,179	\\ \hline \hline
Improvement	&	9.00\%	&	27.4\%	\\

\label{tabUsage}
\end{tabular}
\end{table}

In Table~\ref{tabAccuracy}, we see that the new model is much more {\it accurate} than \Sentinel. Looking at Top1 accuracy and Top3 accuracy we see 9.14 and 14.19 percentage point improvements, respectively, over \Sentinel. These results were statistically significant with $p \ll 0.01$.

When comparing {\it latency} of the new model to \Sentinel, we see major wins in Table~\ref{tabLatency}. The 90th percentile latency decreased from 4.43s to 0.3s (a 14x improvement), and p99 latency decreased from 12.37s to 1.26s (a 9x improvement). This reduces latency largely because our new model uses measures that are stored in a database and does not require blame information to be calculated on the fly. As noted before, we tracked this information in a dashboard outside our experimental framework and were unable to calculate statistical significance. However, from a developer's perspective the latency reduction of 14 times is substantial. 

We also see that diffs in the test condition have more {\it clicks} of the suggested reviewer feature, with a 27.3\% increase in the number of diffs where a suggested  reviewer was added using the recommender reviewer component on the diff review page, see Table~\ref{tabUsage}. 

We did not see any statistically significant change in how long diffs were in review. Importantly, we did not see any statistically significant change in how long reviewers examined the code under review indicating that reviewers are not rubberstamping diffs. Based on these results, \MetaRec has been rolled out to 100\% of engineers at \Meta since early Summer 2022. 

\begin{tcolorbox}

In the A/B test, we saw a Top3 accuracy of 73\%, a 14 point improvement over the previous model. We also saw a 14 times reduction in latency at p90. The recommendations were selected by authors 27\% more often with the new model. These improvements were all statistically significant with $p \ll 0.01$. We did not see any statistically significant changes in the review cycle time or the reviewer viewing time. \MetaRec was rolled out to 100\% of engineers at \Meta in early Summer 2022.

\end{tcolorbox}

\section{Expt. 2: Balancing Reviewer Workload}
\label{secWLResults}

Engineers have different skillsets which will always skew reviewer workload~\cite{rigby2011understanding}, it is still interesting to understand if a recommender that is aware of workload might distribute the load of reviewing and potentially drive down cycle time by suggesting the least busy expert. For example, Microsoft weighted their code reviewer recommender by a candidate's workload and reduced the time it takes for a review~\cite{asthana2019whodo}. A similar experiment at Ericsson~\cite{Strand2020IEricsson} balanced reviewer workload, but was unsuccessful and was rolled back. As a result, we build upon the Microsoft work and we take the score from \MetaRec that is now in production and weight it by the workload of a candidate as described in the design and methods Section~\ref{secDesignWorkload}.

Our goal metric is to reduce the median amount of time spent reviewing the reviewer who made an action on a diff. We expect a decrease in accuracy in Top1 accuracy. However, because we only re-rank reviewers, and still display all five reviewers, the same reviewers will be shown only if the order will have changed.

Our other guardrail metrics are the same as in the previous section. We do not want to see an increase in review cycle time. We want people to remain thorough in reviewing, so the time spent per diff should remain unchanged. We do not want to see fewer overall clicks on our recommendations. 

\subsection{Historical Backtest Results} 

To determine which type of workload is most effective at distributing reviews, we conduct a historical back test on 1.2 million historical diff reviews that were performed in early 2023. The backtest is followed by a production controlled experiment. We backtest three types of workload weights, the number of upcoming meetings in the next seven days, the reviewer activity in the last seven days~\cite{asthana2019whodo}, and time spent reviewing in the last seven days. We calculate each over a seven day period because developer patterns on a particular day tend to vary, but the entire work week tends to be more predictable~\cite{LChen2022FSE-industry}. We also measure the impact on the Top1 and Top3 accuracy. 

The top portion of Table~\ref{tabWorkloadBacktest}, shows the results. We see that the Top1 accuracy decreases dramatically using workload balancing by around -20 percentage points compared to \MetaRec. The workload differences are less consistent, with upcoming meetings performing poorly, and review activity and time spent performing well with respective decreases of -57\% and -73\%. The Top3 accuracy shows similar patterns with time spent having the largest drop in workload. 

Since we are only re-ordering the reviewers suggested by \MetaRec and are still displaying the Top5 recommended reviewers, we decided to use the most aggressive weight metric, time spent reviewing in the last seven days, because it had the largest drop in workload. We implemented the algorithm in production and gated it for an A/B test.

\begin{table*}[h]
\centering
\caption{Backtest comparison of metrics and results from workload balancing in production. We report the percentage point decrease in accuracy and the percentage decrease in workload relative to results from \MetaRec.}
\begin{tabular}{ l | l | r | r | r | r}

{\bf Method}	&	{\bf Workload Metric}	&	{\bf Top1 Accuracy}	&	{\bf Top1 Workload}	&	{\bf Top3 Accuracy}	&	{\bf Top3 Workload}	\\ \hline
Backtest	&	Upcoming Meetings	&	-21.66	&	-1.46	&	-12.66	&	1.55	\\ \hline
Backtest	&	Reviewer Activity	&	-20.61	&	-57.11	&	-13.40	&	-30.03	\\ \hline
Backtest	&	Time Spent	&	-21.37	&	-72.79	&	-20.89	&	-41.14	\\ \hline \hline
Expt. 2	&	Time Spent 	&	-4.80	&	-18.06	&	-5.34	&	-2.84	\\

\end{tabular}
\label{tabWorkloadBacktest}
\end{table*}

\subsection{Results for Expt 2. \MetaRecWL in Production} 

We discussed our experimentation framework and metrics in background Section~\ref{sec:goalDefinition}. To summarize, our goal metric is to reduce the workload of the Top recommended reviewers. Our backtest indicates that we will see a drop in the accuracy. {\it This reduction is the tradeoff for a more evenly distributed workload.} Specifically, we expect the accuracy to drop by no more than -20 percentage points. Our other guardrail metrics remain the same: we do not want to see fewer clicks on our recommendations, we do not want people to spend less time reviewing the diff, and we do not expect to see any change in how long reviews are waiting to be reviewed. The unit in the experiment was the author of a diff, so each author would either have \MetaRec or \MetaRecWL. 
We had 14k authors in each condition for a total of 28k authors. 

\begin{table*}[h]
\centering
\caption{For Expt. 2 on workload balancing, we had \MetaRec as the control and \MetaRecWL as the test. The unit was 28k diff authors. We conducted the experiment in Winter 2023. From our backtest in Table~\ref{tabWorkloadBacktest}, we expected to see a drop in reviewer workload as reviews become more evenly distributed. From the backtest shown in Table~\ref{tabWorkloadBacktest}, we also expected to see a trade-off/drop in accuracy of up to $-21$ pp. The table shows that we did not see as much of a drop in workload, but we also did not see the same drop in accuracy. We did see an unexpected drop in Top 1 Clicks, however, the overall clicks remained constant.}

\begin{tabular}{ l | l | l | c | l }

{\bf Outcome Metric}	&	{\bf Type}	&	{\bf Expectation}	&	{\bf Change}	&	{\bf p-value}	\\ \hline
Workload Top 1	&	Goal	&	Decrease	&	$-18.06$ pp.	&	$p \ll 0.01$	\\ 
Workload Top 3	&	Goal	&	Decrease	&	\NA	&	$p = 0.19$	\\ \hline
Accuracy Top 1	&	Guardrail	&	Decrease must be $ > -21$ pp.	&	$-4.90$ pp.	&	$p \ll 0.01$	\\ 
Accuracy Top 3	&	Guardrail	&	Decrease must be $ > -20$ pp.	&	$-5.34$ pp.	&	$p \ll 0.01$	\\ \hline
Clicks Top 1	&	Guardrail	&	No Decrease	&	$-10$ pp.	&	$p \ll 0.01$	\\ 
All Clicks	&	Guardrail	&	No Decrease	&	\NA	&	$p = 0.09$	\\ \hline
TimeInReview	&	Guardrail	&	No Increase	&	\NA	&	$p = 0.75$	\\ \hline
\eyeballTime	&	Guardrail	&	No Decrease	&	\NA	&	$p = 0.85$	\\

\end{tabular}
\label{tabWorkloadProduction}
\end{table*}

{\bf Results:} We were surprised how different the results from the backtest were from the controlled production experiment, see the last row in Table~\ref{tabWorkloadBacktest}. When the author selected the top recommendation, the accuracy decreased much less than in the backtest -5\% vs -21\% decrease in accuracy and the workload only decreased -18\% vs -73\% in the production vs backtest. The workload drop at Top3 was not statistically significant. 

Our guardrail metric, number of clicks explains the discrepancy, we saw a decrease in selection of the Top1 candidate by 30\% of all diffs to 20\%. Overall, however, the difference in the number of clicks did not change in a statistically significant manner. What this means is that when someone with a low workload was an inappropriate reviewer, the author simply ignored them and moved lower in the list of recommended reviewers. This result also reinforces how important it is to first model expert reviewers with \MetaRec and then simply re-rank.

The remaining guardrail metrics time spent per diff and review cycle time were not changed in a statistically significant manner. This means that we were unable to replicate the Microsoft~\cite{asthana2019whodo} reduction in the time pull-requests spent in review.

\begin{tcolorbox}

When a reasonable candidate with lower workload was available, we saw a large reduction in workload. When an inappropriate candidate was recommended, simply because they had low workload, the author looked down the list for a more experienced developer, and we saw a drop in the top ranked candidate being selected. We did not see any statistically significant change in overall clicks or review cycle time. Since there was a reduction in workload for the Top1 reviewer, \ie a more evenly distributed workload, we replaced \MetaRec with \MetaRecWL and rolled it out to 100\% of diffs at \Meta.

\end{tcolorbox}

\section{The Bystander Effect}
\label{secBystanderResults}

In psychology, the "bystander effect" occurs when a task is assigned to a group of people and leads to a lower likelihood of any individual acting than when an individual is explicitly assigned to a task~\cite{fischer2011bystander}. At \Meta, diff authors can assign teams and groups of developers rather than selecting individual reviewers. In some cases, an author might not mind which developer on a team reviews a diff. Furthermore, \Meta uses a monorepo where developers often make changes that crosscut the codebase and are owned by another development team. The author might not know who the best reviewer is and will instead select a team. Selecting a reviewer group creates the risk of each member thinking, ``someone else will probably review this.” 
We decided to leverage the wins from our reviewer recommendation investments last year to assign a single individual reviewer to diffs where only reviewer groups were assigned in an attempt to reduce the Bystander Effect. 

The code review team decided to take a simple randomization strategy to overcome the bystander effect. \BystanderRnd, is built directly upon \MetaRec, and we simply chose one of the top three reviewers at {\it random}.
Our goal is that reviewers will feel a sense of ownership for diffs where they were assigned as a reviewer, and review it more quickly, thus reducing the \timeInReview metric. We do not want to see regressions in how long a developer views a diff. We do not track latency, clicks, or workload as they are unlikely to be impacted by this experiment. Our experimental unit was 12.5k authors that were evenly divided between the control and test conditions.  

{\bf Results.} \BystanderRnd, was extremely successful and reduced \timeInReview by -11.6\% with $p \ll 0.01$. There was no statistically significant change in the guardrail metric of the amount of time reviewers spent on each diff. We rolled \BystanderRnd out to 100\% of diffs at \Meta where there was only a team or group assigned the review. 

\begin{table*}[h]
\centering
\caption{Expt. 3: the bystander effect. We compared the TeamOnly assignment control to \BystanderRnd test. The experiment involved 12.5k authors evenly divided between the two conditions in Spring 2023. The bystander effect is real, and it was effectively minimized by randomly assigning one of the top three recommended reviewers.}

\begin{tabular}{ l | l | l | c | l }

{\bf Outcome Metric}	&	{\bf Type}	&	{\bf Expectation}	&	{\bf Change}	&	{\bf p-value}	\\ \hline
TimeInReview	&	Goal	&	Decrease	&	-11.6	&	$p \ll .01$	\\ \hline
\eyeballTime	&	Guardrail	&	No Decrease	&	\NA	&	$p = 0.37$	\\

\end{tabular}
\label{tabBystanderResults}
\end{table*}

When a new feature or change is rolled out at \Meta, we post an announcement in our feedback groups. In this case, we had one developer voice concern that explicitly assigning an individual to a diff will take time away from other diffs. To investigate this concern, we looked at all diffs during the time of study and saw no statistically significant change in the time in review per diff for diffs that already had an individual assigned to the diff.  

\begin{tcolorbox}

The bystander effect occurs in code review when only a team is assigned to review a diff instead of an individual. At \Meta we randomly assigned one of the top three experts to review the diff and found a drop of -11.6\% in \timeInReview ($p \ll 0.01$), with no statistically significant regressions in how long reviewers spent per review, \ie \eyeballTime. We rolled \BystanderRnd out all diffs that only had a team assigned, and they will now also have an individual assigned.

\end{tcolorbox}

\section{Threats to Validity}
\label{secThreats}

\subsection{Generalizability}
Drawing general conclusions from empirical studies in software engineering is difficult because any process depends on a potentially large number of relevant context variables. The analyses in the present paper were performed at \Meta, and it is possible that results might not hold true elsewhere. We cannot release our data, even in an anonymized format, because it would violate legally binding employee privacy agreements.
However, the software systems under study cover millions of lines of code and 10s of thousands of developers who are both collocated and working at multiple locations across the world.
We also cover a wide range of domains from user facing social network products and virtual and augmented reality projects to software engineering infrastructure, such as calendar, task, and release engineering tooling. 

There are many reviewer recommenders available in the literature above, and instead of reimplementing these recommenders, we have used the feature categories suggested by Al-Zubaidi \etal~\cite{Al-Zubaidi2020PROMISE}, and included features from each categories that would not increase latency in our model. We have used those that can be extracted from \Meta's history and those that can be calculated in a timely manner. It is possible that smaller organizations or smaller projects can use computationally expensive features, but unfortunately they do not work at our scale.

\subsection{Construct Validity}

The implicit assumption when, using historical data to assess TopN accuracy, is that the person who performed an action on a diff is a good reviewer. In our backtest, we saw that the TopN accuracy can be quite different from the person who actually conducts the review in a controlled experiment in production. Virtually all prior work has assumed that authors would accept the recommendations, our work sheds doubt on historical benchmarks as the gold standard of accuracy. 

In our experiments, we used the outcome measures that are commonly used by the code review team at \Meta. The two most commonly used measures are \timeInReview and \eyeballTime. Most works capture the wall time from when the pull-request is published until it is merged or abandoned. In contrast, at \Meta we only capture the time that a diff is actively being reviewed or is waiting for review measures. Clicks and latency are uncontroversial measures that are simple to calculate. 

Engineers have a variety of tasks and in this work we only backtested the number of meetings, reviewing activity, and the amount of time spent viewing diffs as measures of workload. In the end we selected \eyeballTime, but we look forward to seeing what type of workload is effective at other companies. 

\subsection{Internal Validity}

 
Each of the recommenders presented in this paper has been tested in a production experiment based on goal and guardrail metrics. However, each experiment was run at a different time, with \MetaRec experimentation in the Spring of 2022, \MetaRecWL in Winter of 2023, and \BystanderRnd in Spring of 2023. We cannot report the exact time-frames, but report the number of diffs or diff authors so that the size of the sample can be evaluated by the reader. As part of engineering at \Meta we track metrics across time and did not see any changes in accuracy, \timeInReview, and \eyeballTime outside the experiments that we ran. 

While the accuracy of the recommender is unaffected by collinearity among the features, the relative feature importance can be impacted. For example, we have the feature of the total amount of time a reviewer examined an author’s diff and the time they did this when they were explicitly assigned to review the author’s diffs. These features are definitely correlated and one is even a subset of the other. We kept both features because the engineering team thought that the features were both interesting, and as can be seen in the table, both add predictive power to the model. While the exact feature importance might vary depending on the project or environment, we based our feature selection on categories of features found in the literature and feel that they are a robust representation of features that are useful in reviewer recommendation. 

\subsection{Conclusion Validity}
Conclusion Validity~\cite{wohlin2012experimentation} is defined as  the degree to which conclusions reached (e.g. about relationships between factors) are reasonable within the data collected. We use experiments that randomize treatments, thus eliminating researcher bias. 

We do not explicitly control for factors including the size and complexity of the diff (and of the underlying codebase), author and reviewer expertise, programming language, type of functionality being
implemented, time of the year, and day of the week. Instead we use randomization to help control for observed and unobserved variables, and we also have large sample sizes.

We report all experiments that were conducted and the results for all goal metrics (whether statistically significant or not). The p-values are adjusted for multiple comparisons in case more than one goal metric was used. Finally, since we only compare the distribution of response variables to a single predictor (experimental condition), we minimize the subjectivity due to researchers selecting one versus another subset of predictors for reporting.

\section{Literature and Discussion}
\label{secLitDiscuss}

We interleave our discussion with the literature showing how we advance the state-of-the-art reviewer recommenders.

\subsection{General Reviewer Recommenders}
The first reviewer recommendation system we are aware of was introduced by Balachandran~\etal~\cite{balachandran2013reducing}.
They used authorship of the changed lines in a code review (using \texttt{blame}) to identify who had worked on that code before and suggested a ranked list of this set as potential reviewers.
Similarly, Thongtanunam~\etal~\cite{thongtanunam2015should} proposed \textsc{RevFinder}, a reviewer recommender based on file locations.  \textsc{RevFinder} is able to recommend reviewers for new files based on reviews of files that have similar paths in the filesystem.  The approach was evaluated on over 40,000 code reviews across three OSS projects, and recalls a correct reviewer in the Top10 recommendations 79\% of the time on average. Like \Sentinel these approaches would suggest reviewers who had moved onto other projects because they had edited the files in the past.  

There has been a lot of interest in consuming features beyond file history when building reviewer recommendation models. Rahman~\etal~\cite{rahman2016correct} propose \textsc{CORRECT} an approach to recommend reviewers based on their history across all of GitHub as well as their experience with certain specialized technologies associated with a pull request.
Jiang~\etal~\cite{jiang2017should} examine the impact of various attributes of a pull request on a reviewer recommender, including file similarity, PR text similarity, social relations, and ``activeness'' and time difference.
They find that adding measures of activeness to prior models increases performance considerably.
Zhang~\etal~\cite{Zhang2022UsingLH} built a Socio-Technical graph to capture the social interactions among authors and reviewers. Further, they use Graph Convolution Networks (GCNs) to encode the social interaction information into high-dimensional space, and formulate reviewer recommendation as a link prediction problem. We describe the features that we were able to include in our model in Table~\ref{tableRecFeatures} building upon the  categories suggested by Al-Zubaidi \etal~\cite{Al-Zubaidi2020PROMISE}. 

\subsection{Latency and Performance.}
Some of these recommenders use social network features and other features that require substantial processing. As noted by Microsoft~\cite{asthana2019whodo,Zanjani2016TSE} and Ericsson~\cite{Strand2020IEricsson}, these cannot be used in production systems due to processing delay. We are unaware of any paper that reports results on latency of reviewer recommendations. Interestingly, when we reduced the latency from \Sentinel to \MetaRec we saw that lower latency resulted in authors seeing recommendations earlier before they had decided on who would do the review and many more clicks on recommendations. Future works should provide information on prediction latency. 

\subsection{Reviewer Workload and Recommenders}

Kovalenko \etal~\cite{kovalenko2018does} found that expertise based recommenders tend to suggest reviewers that are always obvious to authors 52\% and 63\% of the time at Microsoft and JetBrains respectively. As a result, authors often clicked on because it is easier to click than type a reviewer's name. Another interesting finding was that authors often found that outdated recommendations were made based on people who use to work on the files. Interviewed developers wanted more information about the current knowledge level of the candidate reviewers as well as how busy, \ie what the workload of the reviewers are. 

Microsoft weighted their code reviewer recommender, \cHRev, by the number of open reviews a candidate, \whodo, had and found between a 14\% and 21\% improvement in the time taken for reviews depending on the project~\cite{asthana2019whodo}. We were unable to replicate this result at \Meta and found no statistically significant change in review cycle time. We did find a decrease in workload of selected reviewers. Unfortunately, Microsoft did not report the change in reviewer workload, so we are unable to make comparisons. 

Al-Zubaidi \etal~\cite{Al-Zubaidi2020PROMISE} treat reviewer recommendation as a multi-objective search problem where they maximize for the chance of participating in review and minimize the shannon entropy, \ie skewness, of the people selected. Their recommender considers review workload, but has low precision ranging from 16\% to 20\% depending on the open source project. They do not use a measure of workload as an outcome, so it is unclear if their approach succeeds in spreading workload. They do not report on latency or performance, and it is unclear if this search-based approach could be used at large scale companies. 

Strand ~\etal~\cite{Strand2020IEricsson} evaluate a workload aware recommender at Ericsson. In a backtest they found that a recommender aware of the file change history and the level of recent workload activity was able to make recommendations with an MRR of 0.37 or an average rank of 2.7. Despite these reasonable recommendations, after manually evaluating the performance of the recommender on 47 changes they determined that it did not reduce lead time or workload for code review and they decided to not roll-out the recommender. Like Ericsson, we did not find a change in cycle time, however, we did find changes in the workload of the reviewer.

Hajari ~\etal~\cite{Hajari2024TSE} used the Gini coefficient to capture how unequal workload is and on open source projects found a near perfect Pareto principle with the top 20\% of reviewers doing 80\% of the reviews. The authors simulate the change in Gini over multiple years of data and conclude that a workload aware recommender similar to those used by Microsoft~\cite{asthana2019whodo} can reduce workload. Our work threatens the conclusions of this paper because workload balanced recommendations can be ignored with the ``normal'' experts still doing most of the work. It is interesting future work to see if the Gini co-efficient changes over the period of multiple years when a workload balanced recommender is released in production.

Murphy-Hill \etal~\cite{MurphyHill2023CSCW}, examine workload at Google and discover a systemic imbalance in gender and code review selection. Authors manually select men more often than women as reviewers. Since recommenders are trained on prior data, the recommenders contain this bias. 
In our work, we do not explicitly factor gender into our recommender, however, we re-rank reviewers based on workload, so if women are assigned fewer reviewers, they will automatically be assigned more reviews as workload is balanced. We also use random in our bystander experiment, which Murphy-Hill \etal finds balances the gender of recommended reviewers.

\section{Conclusion}
\label{secConclusion}

At \Meta we make the following contributions related to accuracy and latency, workload, and the bystander effect:

\begin{enumerate}

\item \MetaRec: In the A/B test, we saw a Top3 accuracy of 73\%, a 14 point improvement over the previous model. We also saw a 14 times reduction in latency at p90. The recommendations were selected by authors 27\% more often with the new model. We did not see any statistically significant changes in the review cycle time or the reviewer viewing time. \MetaRec has been rolled out to 100\% of engineers at \Meta.

\item \MetaRecWL: When a reasonable candidate with lower workload was available we saw a large reduction in workload. When an inappropriate candidate was recommended, simply because they had low workload, the author looked down the list for a more experienced developer, and we saw a drop in the top ranked candidate being selected. We did not see any statistically significant change in review cycle time or latency. As a result, we decided to only rerank the top three reviewers, and rolled \MetaRecWL out to 100\% at \Meta.

\item \BystanderRnd: The bystander effect occurs in code review when only a team is assigned to review a diff instead of an individual. At \Meta we randomly assigned one of the top three experts to review the diff and found a drop of -11.6\% in \timeInReview with no regressions in our guardrail metrics. We rolled \BystanderRnd out to 100\% of engineers.

\end{enumerate}

Our contributions to the general field of software engineering are as follows: 

\begin{enumerate}

\item We updated Al-Zubaidi \etal's~\cite{Al-Zubaidi2020PROMISE} 2020 literature review. We developed a recommender based on measures from each of their categories. Interestingly, we found that reviewer relationships, \ie how often a reviewer had reviewed an author in the past, was the strongest predictor. Expertise was less important. 

\item We noted a large discrepancy between backtesting (the typical way code review tools are evaluated in research literature) and A/B test. While, obviously, not in all instances A/B test can be performed, future studies may discover ways to do backtesting in a way that is more reflective of what would be obtained via A/B test. In many companies code review is implemented as a service, so A/B test is feasible given that privacy and other ethical considerations are adhered to.

\item We described a set of goal and guardrail metrics that can be used in code reviewer recommendation. We feel that \eyeballTime is particularly important because it allows researchers to ensure that reviewers are not rubberstamping reviews without thoroughly considering the code. The metric can be easily captured using tools that capture the time spent on a webpage.

\item We are unaware of any works that reported on the impact of latency on reviewer recommendation. When our slowest 10\% of recommendations (the 90th percentile recommendation time) went from 4.4 seconds to 300ms, we saw a 27\% increase in the number of recommendations that were clicked on. We suggest that future work should report how long it takes to generate a recommendation, if the 90th percentile is above 300ms, then it is likely that the suggestion will load after the author has already selected other reviewers. 

\item Although prior works attempted to balance reviewer workload, these works did not capture the change in workload, instead measuring other variables like \timeInReview. We suggested a simple measure of workload and showed that some degree of workload balancing can work. We hope that Microsoft, Ericsson, and other companies will report on the impact of workload balancing. 

\item We are unaware of any works that quantify the impact of assigning a team rather than an individual. We were impressed how a simple random assignment of one of top recommended reviewers could effectively mitigate the bystander effect. This finding has implications beyond software firms and may impact how tasks are assigned generally in organizations. We look forward to future work on task assignments.

\end{enumerate}

\bibliographystyle{ACM-Reference-Format}
\bibliography{ref.bib}

\end{document}